\documentclass[usenatbib,onecolumn]{mn2e}
\usepackage{amsfonts}
\usepackage{amsmath}
\usepackage{graphicx}
\usepackage{natbib}
\usepackage{amssymb}

\def\apjl{Astrophys.\ J.\ Lett.}
\def\mnras{Mon.\ Not.\ R.\ Astron.\ Soc.}

\def\aap{Astron.\ Astrophys.}
\def\apj{Astrophys.\ J.}

\def\apjs{Astrophys.\ J. Supp.}

\def\prd{Phys.\ Rev.\ D}

\def\physrep{Phys. Rep.}

\title[Initial shear field in $f_{nl}$ models]
      {The initial shear field in models with primordial local 
       non-Gaussianity and implications for halo and void abundances}

\author[T. Y. Lam, R. K. Sheth \& V. Desjacques]
 {Tsz Yan Lam$^{1,2}$\thanks{E-mail:tylam@sas.upenn.edu, shethrk@physics.upenn.edu,
 dvince@physik.uzh.ch}, 
 Ravi K. Sheth$^1$\footnotemark[1] \& Vincent Desjacques$^3$\footnotemark[1]\\
 $^1$ Center for Particle Cosmology, University of Pennsylvania, 
 209 S. 33rd Street, Philadelphia, PA 19104, USA 
 \\
 $^2$ Institute for the Physics and Mathematics of the Universe, University of Tokyo, Chiba 277-8582, Japan
 \\
 $^3$ Institute for Theoretical Physics, University of Z\"urich, 
 Winterthurerstrasse 190, CH-8057 Z\"urich, Switzerland}

\newcommand{\bm}[1]{{\mbox{\boldmath $#1$}}}

\begin{document}
\pagerange{\pageref{firstpage}--\pageref{lastpage}}

\maketitle

\label{firstpage}

\begin{abstract}
We generalize Doroshkevich's celebrated formulae for the eigenvalues 
of the initial shear field associated with Gaussian statistics to 
the local non-Gaussian $f_{nl}$ model.  This is possible because, to 
at least second order in $f_{nl}$, distributions at fixed overdensity 
are unchanged from the case $f_{nl}=0$.  
We use this generalization to estimate the effect of $f_{nl}\ne 0$ 
on the abundance of virialized halos.  Halo abundances are expected 
to be related to the probability that a certain quantity in the 
initial fluctuation field exceeds a threshold value, and we study 
two choices for this variable:  it can either be the sum of 
the eigenvalues of the initial deformation tensor (the initial 
overdensity), or its smallest eigenvalue.  
The approach based on a critical overdensity yields results which 
are in excellent agreement with numerical measurements.  
We then use these same methods to develop approximations describing
the sensitivity of void abundances on $f_{nl}$.
While a positive $f_{nl}$ produces more extremely 
massive halos, it makes fewer extremely large voids. Its effect thus
is qualitatively different from a simple rescaling of the normalisation
of the density fluctuation field $\sigma_8$. Therefore, void 
abundances furnish complementary information to cluster abundances, and 
a joint comparison of both might provide interesting constraints on
primordial non-Gaussianity.  
\end{abstract}


\section{Introduction}
Detections of non-gaussianity can discriminate between different
inflation models \cite[e.g.][]{jm03}.   The local $f_{nl}$ model,
where the primordial  perturbation potential is
\begin{equation}
 \Phi = \phi + f_{nl}(\phi^2 - \langle \phi^2\rangle),
 \label{eqn:fnl}
\end{equation}
where $\phi$ is a Gaussian potential field and $f_{nl}$ is a scalar,
has been the subject of much recent study \citep[e.g.,][and references
therein]{bko08,kp08,st08}.   Constraints on this model tend to be of
two types -- from the CMB  \citep{cmb5yr,hikageetal08,yw08,mhlm08,rsphmfnl}
and from large scale structures in the Universe
\citep{kst99,mvj00,ssz04,sk07,is07,fnlverde,dalaletal08,mv08,cvm08,at08,shshp08,mcdonald08,tkm08,slosar08,grossi08,kvj08,pphfnl08,fnlvincent,lamshethfnl,grossinfm}.

The initial shear field is expected to play an important role in  the
formation of large scale structures
\citep{zel70,bm96,ls98,smt01,ptreview,d07}.   The main goal of the present
work is to show that much of the  machinery developed for the study of
structure formation from  Gaussian initial conditions can be carried
over, with minor modifications,  to the study of $f_{nl}$ models.  We
do so by showing how to generalize  Doroshkevich's celebrated formulae
for the eigenvalues of the initial  shear field associated with
Gaussian statistics \citep{grf} to the local non-Gaussian $f_{nl}$
model.

We then study how the abundance of virialized dark matter halos
depends on $f_{nl}$. This study focuses on two problems:  one is the
physical  model for halo formation, and the other is the statistical
problem  of how this collapse model is used to estimate the abundance
of  such collapsed objects.  We use the approach pioneered by
\cite{ps74}  and refined by \cite{bcek91} to address the statistical
problem \cite[see][for recent extentions which treat the $f_{nl}\ne 0$
case]{mr09b, lsfnlhalo}.   We study two different models for the
physics of halo formation:   one in which halos form from sufficiently
overdense regions in the  initial fluctuation field
\citep{ps74,smt01}, and another in which  the criterion for halo
formation is that all three eigenvalues of  the initial deformation
tensor exceed a certain value \cite[e.g.][]{ls98}.   In the former, we
use the triaxial collapse model of \cite{bm96} to  estimate this
critical overdensity \cite[following][]{smt01}.   If our way of
estimating halo abundances are reliable, then comparison  with
simulations run for a range of $f_{nl}$ values provides a novel  way
to study the physics of gravitational clustering.

Section~\ref{section:def} provides explicit expressions for the 
initial eigenvalue distribution, and for the initial distribution 
of the variables which arise naturally in the context of triaxial 
collapse models.  These are used, in Section~\ref{section:mf}, to 
estimate how the mass function of virialized objects is modified 
when $f_{nl}\ne 0$.  This Section also shows the result of comparing 
these estimates with measurements in simulations.  
With some care \cite[e.g.][]{sw04}, much of the analysis can be 
carried over straightforwardly to study void abundances; this is the 
subject of Section~\ref{section:vf}.  
A final section summarizes our results.
An Appendix describes an alternative estimate of halo abundances 
which is logically consistent with previous work, but which does not 
reproduce the $f_{nl}$ dependence seen in simulations.


\section{The local non-gaussian model} \label{section:def}
We are interested in models where the primordial perturbation 
potential is given by equation~(\ref{eqn:fnl}).
We will use $P_{\phi}(k)$ to represent the power spectrum of $\phi$; 
in what follows we will set $P_{\phi}(k)= Ak^{n_s-4}$, where 
$n_s\approx 1$, and $A$ is a normalization constant that is fixed by 
requiring that the rms fluctuation in the associated non-Gaussian 
initial density field (which we will define shortly) has value $\sigma_8$.
The power spectrum and bispectrum of the $\Phi$ field are 
\begin{align}
 P_{\Phi}(k) & = P_{\phi}(k) + \frac{2f_{nl}^2}{(2\pi)^3}
   \int {\rm d}\, {\bm q} \left[P_{\phi}(q)P_{\phi}(|{\bm k} - {\bm q}|)
  - P_{\phi}(k)P_{\phi}(q) - P_{\phi}(k)P_{\phi}(|{\bm k} - {\bm q}|)\right],\\
 B_{\Phi}(k_1,k_2,k_{12}) & \equiv 
    2f_{nl}\,\left[ P_{\phi}(k_1)P_{\phi}(k_2) + 
    {\rm cyclic} \right] + \mathcal{O}(f_{nl}^3) \label{eqn:bkphi}
\end{align}
\citep{ssz04}.

\subsection{The shear or deformation tensor}
Define ${\bm D}$ as the real, symmetric $3\times 3$ tensor whose 
components are proportional to the second order derivatives of the 
potential $\Phi$: 
\begin{equation}
 \Phi_{ij} \equiv \phi_{ij} + 2f_{nl}(\phi_i\phi_j + \phi\phi_{ij}),
\end{equation}
where $\phi_i = \partial_i \phi$ and 
      $\phi_{ij} = \partial_i\partial_j \phi$. 
We will sometimes refer to ${\bm D}$ as the shear field associated 
with the potential $\Phi$.  
Correlations between the $\Phi_{ij}$ will be very useful in what 
follows.  These depend on the correlations between $\phi$ and its 
derivatives but, because $\phi$ is Gaussian, they can be computed 
easily.  Doing so shows that the six components of ${\bm D}$ are 
not independent:  although the three off-diagonal terms are not 
correlated with the others, the three diagonal terms are.  
However, if we set 
\begin{equation}
 x = \sum_i \Phi_{ii}, \qquad 
 y = \frac{1}{2}(\Phi_{11} - \Phi_{22}), \qquad
 z = \frac{1}{2}(\Phi_{11} + \Phi_{22} -2 \Phi_{33}),
\end{equation}
then these three parameters, combined with
 $({\Phi_{12},\Phi_{23},\Phi_{31}})$, 
form a new set of six independent components \citep{bbks86}. 
When $f_{nl}=0$, then each of these is an independent Gaussian 
random field.  

Most of the complication in $f_{nl}$ models arises from the fact 
that we are almost always interested in spatially smoothed quantities.  
Fortunately, smoothing is a linear operation, and the new variables 
$x,y,z$ are just linear combinations of the elements of ${\bm D}$.  
Hence, if $W(kR)$ denotes the Fourier transform of the smoothing 
window of scale $R$, to second order in $f_{nl}$, 
\begin{align}
 \langle x^2 \rangle & = \sigma^2, 
 & \langle y^2 \rangle & = \frac{\sigma^2}{15}, 
 & \langle z^2 \rangle & = \frac{\sigma^2}{5}, 
 & \langle\Phi_{ij}^2\rangle_{i \neq j} & = \frac{\sigma^2}{15} \\
 \langle x^3 \rangle & = 2f_{nl} \gamma^3 
 & \langle y^3 \rangle & = 0 
 & \langle z^3\rangle  & = 0 
 & \langle \Phi_{ij}^3\rangle_{i\neq j} & = 0,
 \label{xyzSmoothed}
\end{align}
where 
\begin{align}
 \sigma^2 & = \frac{1}{(2\pi)^3}\int \frac{{\rm d} k}{k} 
       4\pi\, k^7 M^2(k)\,P_{\Phi}(k)\, W^2(kR), \label{eqn:variance} \\
 \gamma^3 & = \frac{2}{(2\pi)^4}
               \int  \frac{{\rm d} k_1}{k_1} k_1^5 M(k_1) W(k_1R)
               \int  \frac{{\rm d} k_2}{k_2} k_2^5 M(k_2) W(k_2R)
               \int {\rm d}\mu_{12}\, k_{12}^2\, M(k_{12}) W(k_{12}R) 
               \frac{B_{\Phi}(k_1,k_2,k_{12})}{2f_{nl}} \label{eqn:skewness}
\end{align}
and $M(k) \equiv (3D(z)c^2)/(5\Omega_m H_0^2)\,T(k)$, where $T(k)$ 
is the CDM transfer function and $D(z)$ is the linear growth function.  
In what follows, the quantity
 $\sigma S_3\equiv \sigma\,\langle x^3\rangle/\langle x^2\rangle^2$, 
will play an important role, because it represents the leading order 
contribution to the non-Gaussianity (note that it is proportional to 
$f_{nl}$).  Appendix~A in \cite{lsfnlhalo} provides a useful fitting 
formula for this quantity.

\subsection{Joint distribution of eigenvalues}
Equation~(\ref{xyzSmoothed}) shows that, to first order in $f_{nl}$, 
five of the six parameters have zero skewness, so, to first order 
in $f_{nl}$, all but $x$ are drawn from Gaussian distributions.  
However, $x$ is the trace of ${\bm D}$, so it is the linear theory 
overdensity $\delta$.  If $p(\delta|R)$ denotes the distribution of 
$\delta$ when smoothed on scale $R$, then the fact that the other 
variables have the same distribution as in the case $f_{nl}=0$ allows 
one to provide an excellent analytic approximation to the joint 
distribution of the eigenvalues.  Namely,
\begin{equation}
 p({\bm \lambda}|R) = p(\delta|R)\,
        \frac{3^4/4}{\Gamma(5/2)}
        \left(\frac{5}{2\sigma^2}\right)^{5/2}
      \,\exp\left(-\frac{5\delta^2}{2\sigma^2} + \frac{15I}{2\sigma^2}\right)
      \,(\lambda_1-\lambda_2)(\lambda_2-\lambda_3)(\lambda_1-\lambda_3) ,
 \label{eqn:pdflambdas}
\end{equation}
where $\delta\equiv \lambda_1+\lambda_2+\lambda_3$, 
 $I \equiv \lambda_1\lambda_2 + \lambda_1\lambda_3 + \lambda_2\lambda_3$, 
and recall that $\sigma$ is a function of $R$.  
Our convention is $\lambda_1 \geq \lambda_2 \geq \lambda_3$.  
This has the same form as Doroshkevich's (1970) formula for Gaussian 
fields; the only difference is that here $p(\delta|R)$ is not Gaussian
(we provide an expression for it in equation~\ref{eqn:pdfdeltal} below).  

The fundamental reason why this works is that Doroshkevich's formula 
is actually the product of two independent distributions, one of 
$\delta$, and the other of a quantity which is a combination of the 
five other independent elements of the deformation tensor 
\citep[e.g.][]{st02}.  
Since the distribution of each of these other elements is unchanged
from the Gaussian case (we just showed that they all have zero
skewness), this second distribution is unchanged from that of the
Gaussian case -- the only change is the distribution of $\delta$.   In
fact, this holds for any local mapping $\Phi=f(\phi)$  (and is also
true for the alignment of the principal axes, see \cite{ds08}).

\subsection{Distributions at fixed $\delta$}
One consequence of this is that distributions at fixed $\delta$ 
are the same as in the Gaussian case.  For example, 
\begin{equation}
 p(\lambda_i,\lambda_j|\delta) = 
     \frac{3^4/4}{\Gamma(5/2)}\left(\frac{5}{2\sigma^2}\right)^{5/2}
   \,\exp\left(-\frac{5\delta^2}{2\sigma^2} + \frac{15I_{ij}}{2\sigma^2}\right)
   \,(\lambda_i-\lambda_j)(\lambda_i + 2\lambda_j-\delta)
     (2\lambda_i+\lambda_j - \delta) 
   = p_0(\lambda_i,\lambda_j|\delta) , 
\end{equation}
where $i\ne j$ can take values from 1 to 3, 
 $I_{ij}=\lambda_i\lambda_j + (\lambda_i+\lambda_j)(\delta-\lambda_i-\lambda_j)
    = \delta\,(\lambda_i+\lambda_j)
            - (\lambda_i^2 + \lambda_i\lambda_j + \lambda_j^2)$, 
and the subscript $0$ indicates the distribution associated with 
Gaussian initial conditions, for which $f_{nl}=0$.  
Integrating over one of the eigenvalues in the expression above, 
e.g., $\lambda_i$, yields an expression for the distribution 
of the other at fixed $\delta$.  Clearly, such expressions will 
also be the same as in the Gaussian case.  Hence, 
\begin{equation}
 p(\lambda_j|\delta) = p_0(\lambda_j|\delta)
\end{equation}
for $j=1,2,3$.

\subsection{Edgeworth approximation for $p(\delta|R)$}
Because we are interested in small departures from Gaussianity, 
the Edgeworth expansion provides a convenient form for the 
distribution of $\delta$:  
\begin{equation}
 p(\delta|R)\, {\rm d}\delta \approx \left[
 1 + \frac{\sigma(R)S_3(R)}{6}H_3\left(\frac{\delta}{\sigma(R)}\right)\right]\ 
  \frac{e^{-\delta^2/2\sigma^2(R)}}{\sqrt{2\pi}\sigma(R)}\,{\rm d}\delta
  = \left[1 + \frac{\sigma S_3}{6}H_3(\nu)\right]\,
    p_0(\delta|R)\, {\rm d}\delta,
\label{eqn:pdfdeltal}
\end{equation}
where $\sigma(R)$ is given by equation~(\ref{eqn:variance}), 
$\sigma S_3 \equiv \langle x^3\rangle/\langle x^2\rangle^{3/2} 
                 = 2f_{nl}\gamma^3/\sigma^3$,
and $H_3(\nu) \equiv \nu(\nu^2 - 3)$ with $\nu\equiv \delta/\sigma(R)$.  
(See \cite{fnlverde} and \cite{lamshethfnl} for previous work with the 
Edgeworth expansion in the context of $f_{nl}$ models.)  
The final equality writes $p$ as a correction factor times the 
Gaussian distribution $p_0$ to highlight the fact that 
\begin{equation}
 p({\bm \lambda}|R) = 
 \left[1 + \frac{\sigma S_3}{6}H_3\left(\delta/\sigma\right)\right]
  \,p_0({\bm \lambda}|R)\,
\end{equation}
where $p_0({\bm\lambda}|R)$ is Doroshkevich's formula.

\subsection{Distribution of eigenvalues}
Replacing $p(\delta|R)$ in eq.(\ref{eqn:pdflambdas}) by its Edgeworth expansion
and integrating over two of the three eigenvalues, we can write
\begin{equation}
 p(\lambda_i) = p_0(\lambda_i) + \frac{\sigma S_3}{6}\,\Delta p(\lambda_i)
\end{equation}
where 
\begin{eqnarray}
 p_0(\lambda_{1}) & = &
 \frac{\sqrt{5}}{12\pi\sigma}\Bigg{\{} 20\frac{\lambda_1}{\sigma}
 \exp\bigg{(}-\frac{9\lambda_{1}^2}{2\sigma^2}\bigg{)}
 - \sqrt{2\pi}\exp\bigg{(}-\frac{5\lambda_{1}^2}{2\sigma^2}\bigg{)}
 \bigg{(}1-20\frac{\lambda_{1}^2}{\sigma^2}\bigg{)} 
 \left(1 + {\rm erf}\bigg{(}\sqrt{2}\frac{\lambda_{1}}{\sigma}\bigg{)}\right)
\nonumber \\
  &&\qquad\qquad + 3\sqrt{3\pi} 
 \exp\bigg{(}-\frac{15\lambda_{1}^2}{4\sigma^2}\bigg{)}
 \left(1 + {\rm erf}\bigg{(}\frac{\sqrt{3}\lambda_{1}}{2\sigma}\bigg{)}\right)
 \Bigg{\}}
  \nonumber  \\
  p_0(\lambda_{2}) & = & 
 \frac{\sqrt{15}}{2\sqrt{\pi}\sigma}\exp\bigg{(}-\frac{15\lambda_{2}^2}
 {4\sigma^2}\bigg{)}, \\ \nonumber \\
 p_0(\lambda_{3}) & = &
-\frac{\sqrt{5}}{12\pi\sigma}\Bigg{\{} 20\frac{\lambda_3}{\sigma}
\exp\bigg{(}-\frac{9\lambda_{3}^2}{2\sigma^2}\bigg{)}
+ \sqrt{2\pi}\exp\bigg{(}-\frac{5\lambda_{3}^2}{2\sigma^2}\bigg{)}
{\rm erfc}\bigg{(}\sqrt{2}\frac{\lambda_{3}}{\sigma}\bigg{)}\bigg{(}1-20
\frac{\lambda_{3}^2}{\sigma^2}\bigg{)} \nonumber \\
&&\qquad\qquad - 3\sqrt{3\pi}
  \exp\bigg{(}-\frac{15\lambda_{3}^2}{4\sigma^2}\bigg{)}
  {\rm erfc}\bigg{(}\frac{\sqrt{3}\lambda_{3}}{2\sigma}\bigg{)}\Bigg{\}}. 
\end{eqnarray}
and 
\begin{align}
 \Delta p(\lambda_1) = & \frac{\sqrt{5}}{12\pi\sigma}
  \left[\frac{25}{27} \left(8 - \frac{435\lambda_1^2}{4\sigma^2} 
                              + 100\frac{\lambda_1^4}{\sigma^4}\right)
                      \exp\left(-\frac{9\lambda_1^2}{2\sigma^2}\right) 
    + \sqrt{2\pi}\frac{25\lambda_1}{27\sigma}
       \left(51-185\frac{\lambda_1^2}{\sigma^2} 
             + 100\frac{\lambda_1^4}{\sigma^4}\right)
       \exp\left(-\frac{5\lambda_1^2}{2\sigma^2}\right)
       \left(1+{\rm erf}\left(\sqrt{2}\frac{\lambda_1}{\sigma}\right)\right)
 \right. \nonumber \\
& \left. \qquad\qquad + 3\sqrt{3\pi} \frac{25\lambda_1}{4\sigma}
  \left(\frac{5\lambda_1^2}{2\sigma^2} - 1\right)
  \exp\left(-\frac{15\lambda_1^2}{4\sigma^2}\right)
  \left(1+{\rm erf}\left(\frac{\sqrt{3}\lambda_1}{2\sigma}\right)\right) \right],\\
 \Delta p(\lambda_2) = & \left(\frac{5}{6}\right)^{3/2}
                         \frac{\sqrt{15}\lambda_2}{\sqrt{2}\sigma}
       \left(\frac{15\lambda_2^2}{2\sigma^2} - 3\right)\,p_0(\lambda_2)
       = \left(\frac{5}{6}\right)^{3/2}\,
          H_3\left(\frac{\sqrt{15}\lambda_2}{\sqrt{2}\sigma}\right)\,
          p_0(\lambda_2),\\
\Delta p(\lambda_3) = & \frac{\sqrt{5}}{12\pi\sigma} 
  \left[-\frac{25}{27}\left(8 - \frac{435}{4}\frac{\lambda_3^2}{\sigma^2} 
                             + 100\frac{\lambda_3^4}{\sigma^4}\right)
     \exp\left(-\frac{9\lambda_3^2}{2\sigma^2}\right) 
   + \sqrt{2\pi}\frac{25\lambda_3}{27\sigma}
                \left(51 - 185\frac{\lambda_3^2}{\sigma^2}
                         + 100\frac{\lambda_3^4}{\sigma^4}\right)
                \exp\left(-\frac{5\lambda_3^2}{2\sigma^2}\right)
 {\rm erfc}\left(\sqrt{2}\frac{\lambda_3}{\sigma}\right) \right. \nonumber \\
& \left. \qquad\qquad + 3\sqrt{3\pi} \frac{25\lambda_3}{4\sigma}
  \left(\frac{5\lambda_3^2}{2\sigma^2}-1\right)
  \exp\left(-\frac{15\lambda_3^2}{4\sigma^2}\right)
  {\rm erfc}\left(\frac{\sqrt{3}\lambda_3}{2\sigma}\right) \right].
\end{align}
These results should be easily extended to the distribution of shear eigenvalues 
at multiple points (e.g., \cite{ds08}).

\begin{figure}
\centering
 \includegraphics[width=0.9\linewidth]{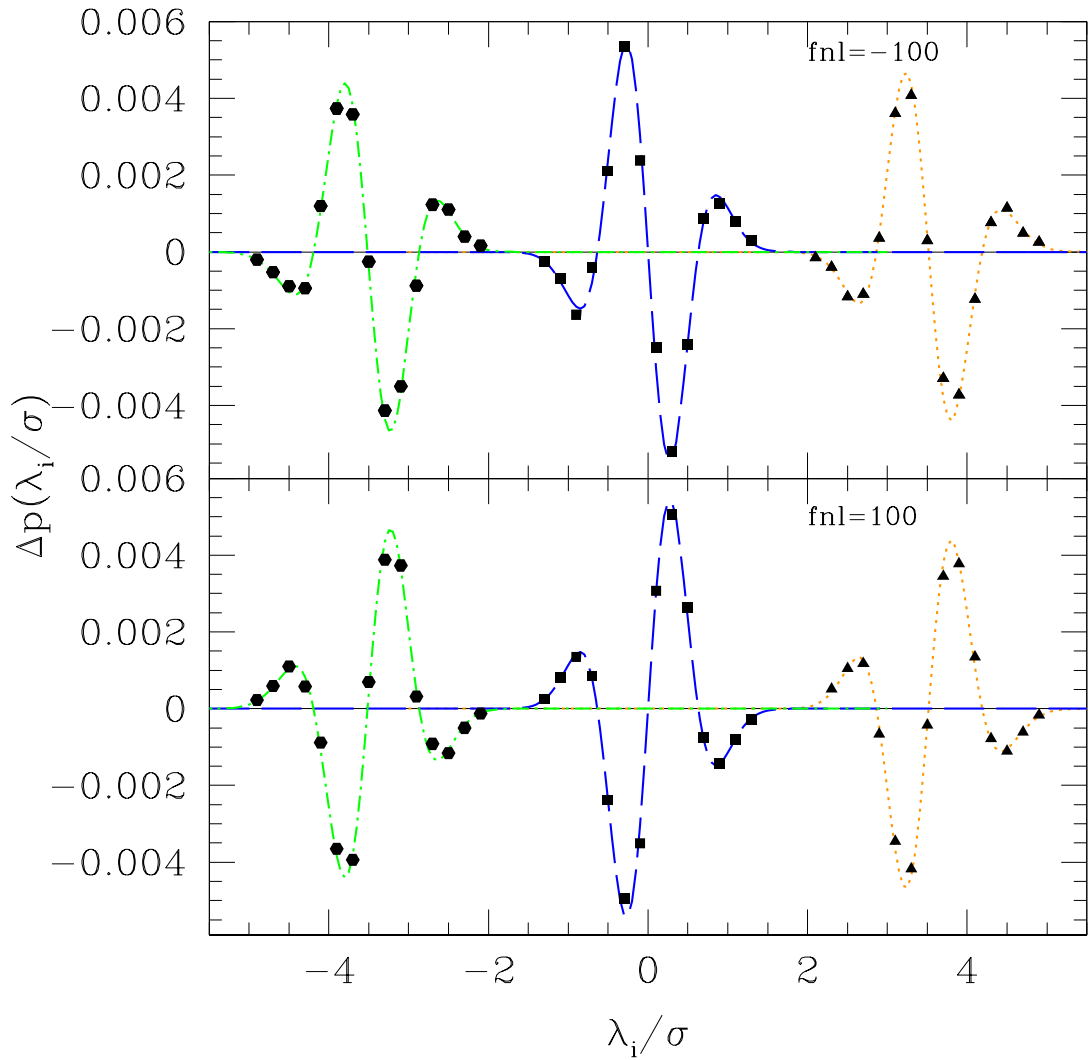}
 \caption{Difference between initial distributions in the 
         $f_{nl}$ and Gaussian models for $f_{nl}=100$ (lower panels) 
         and $f_{nl}=-100$ (upper panels) when the smoothing scale is 
         $1h^{-1}$Mpc. 
	 In both panels, the right and left plots show $p(\lambda_1/\sigma)$
	 (orange curve, triangle symbols) and $p(\lambda_3/\sigma)$
	 (green curve, hexagonal symbols), respectively. For clarity, these
	 curves have been shifted to the right and to the left by 3. The 
	 middle plot shows $P(\lambda_2/\sigma)$ (blue curve, square symbols).
          }
 \label{fig:initpdfs}
\end{figure}

\begin{figure}
\centering
 \includegraphics[width=0.9\linewidth]{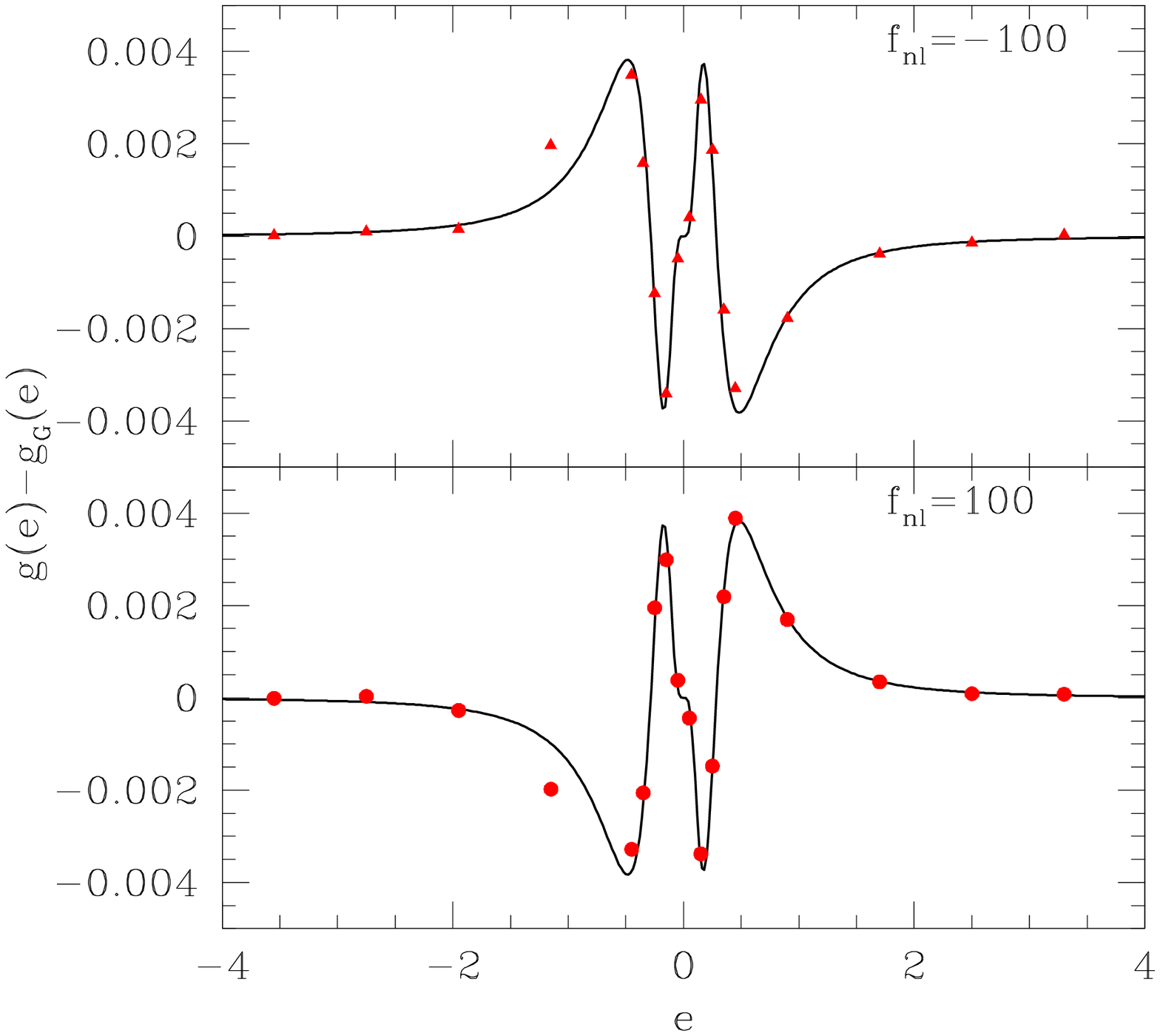}
 \caption{Difference between initial distributions of ellipticity $e$ in the 
         $f_{nl}$ and Gaussian models for $f_{nl}=100$ (lower panels) 
         and $f_{nl}=-100$ (upper panels) when the smoothing scale is 
         $1h^{-1}$Mpc. }
 \label{fig:initge}
\end{figure}

\subsection{Distribution of $\delta$, $e$ and $p$}
In addition to the individual probability distriburions of the three
eigenvalues, we can also derive expressions for the  quantities of
most interest in the ellipsoidal collapse model.   These are the
ellipticity $e$ and prolateness $p$, where
\begin{equation}
 e = \frac{\lambda_1 - \lambda_3}{2\delta}, \qquad {\rm and}\qquad 
 p = \frac{\lambda_1+\lambda_3 - 2\lambda_2}{2\delta}
   = \frac{1}{2} - \frac{3\lambda_2}{2\delta}
   = e - \frac{\lambda_2 - \lambda_3}{\delta}.
\end{equation}
The discussion above means that the distribution of $e$ and $p$ at 
fixed $\delta$ are unchanged from the Gaussian case:
\begin{equation}
 g(e,p|\delta) = \frac{1125}{\sqrt{10\pi}}\,e\,(e^2-p^2)\,
                   \left(\frac{\delta}{\sigma}\right)^5\,
                   {\rm e}^{-(5/2)(\delta/\sigma)^2(3e^2+p^2)}.
 \label{gepd}
\end{equation}
However, the distribution of ellipticity is changed:  
\begin{equation}
g(e) = \int {\rm d}p \int\, {\rm d}\delta\, g(e,p|\delta)\,p(\delta) 
     \equiv g_0(e) + \frac{\sigma S_3}{6}\,\Delta g(e), 
 \label{ge}
\end{equation}
where 
\begin{equation}
 g_0(e) = \frac{45e}{\pi}\frac{1}{(1+20e^2)(1+15e^2)^{5/2}}\left[\sqrt{5}e(1+30e^2)\sqrt{1+15e^2} - (1+20e^2)\arctan\left(\frac{\sqrt{5}e}{\sqrt{1+15e^2}}\right)\right],
\end{equation}
and 
\begin{equation}
 \Delta g(e) = \frac{-45000}{\sqrt{10\pi}}\frac{e}{|e|}\,e^4
   \left[ \frac{4725e^6 + 90e^4-26e^2-1}{(1+15e^2)^4(1+20e^2)^{5/2}}\right].
\end{equation}

To check these expressions, we have used a Monte Carlo method to generate 
 $(x,y,z,\Phi_{12},\Phi_{23},\Phi_{31})$ 
(all but $x$ are drawn from Gaussian distributions).  
We then solve the eigenvalue problem (by solving the associated cubic 
equation) to obtain  $(\lambda_1,\lambda_2,\lambda_3)$ and hence 
$(\delta_l,e,p)$.  We then compute the distributions of the eigenvalues, 
and of $\delta$, $e$ and $p$, and compare them with the associated 
quantities when $f_{nl}=0$.  
The symbols in Figure~\ref{fig:initpdfs} show our Monte Carlo 
results when the smoothing scale is $1h^{-1}$Mpc, and the smooth 
curves show the analytic formulae derived above.  Notice the reflection symmetries
in Figures~\ref{fig:initpdfs} and \ref{fig:initge} 
of the opposite sign of $f_{nl}$: it is due to the fact switching the sign of $f_{nl}$ only 
changes the sign of $\sigma S_3/6$ without modifying other terms in $\Delta p(\lambda_i)$
and $\Delta g(e)$.

\section{Halo abundances}\label{section:mf}
\cite{ps74} (hereafter PS) argued that the abundance of collapsed 
virialized halos may be estimated from the statistics of the initial 
fluctuation field.  They used the assumption that halos form from a 
spherical collapse to argue that such objects started out as 
sufficiently overdense regions in the initial fluctuation field.  
They then used Gaussian statistics to estimate collapsed halo 
abundances.  Although the way in which they used Gaussian 
statistics to make the estimate is flawed, 
\cite{ls98} (hereafter LS) suggested that one might be able to 
provide a better estimate of the abundance of collapsed halos by 
repeating the PS argument, but changing the collapse model to allow 
halo formation to be nonspherical.  In particular, they suggested 
that one should identify halos with regions in the initial field 
where all three eigenvalues were greater than some critical value, 
$\lambda_c$.  
We will use the analysis above to show how this estimate of 
the halo mass function depends on $f_{nl}$.  

\subsection{If halo formation depends on the initial overdensity exceeding a critical value}
The PS-like estimate of the mass fraction in halos above mass $M$ is 
\begin{equation}
 F(>M) = F(<\sigma(R)) 
       = \int_{\delta_c}^\infty {\rm d}\delta\, p(\delta|R)\,
       = \int_{\delta_c}^\infty {\rm d}\delta\, 
             \left[1 + \frac{\sigma S_3}{6}H_3\right]\, p_0(\delta|R),
 \label{eqn:PS}
\end{equation}
where $R = (3M/4\pi\bar\rho)^{1/3}$, so $\sigma(R)$ is actually a 
function of $M$, and $\delta_c$ is the critical density required for 
collapse in the spherical model.  So,  
\begin{equation}
 \frac{\partial F}{\partial\sigma} 
       =  \frac{\partial F_0}{\partial\sigma} + 
     \frac{\partial (\sigma S_3/6)}{\partial\sigma} 
     (\nu_c - \nu_c^{-1})\, \nu_c\,p_0(\nu_c)
     - \frac{\sigma S_3}{6}
        \frac{\partial\ln\nu_c}{\partial\sigma} 
        \nu_c\,p_0(\nu_c)\,H_3(\nu_c),
\end{equation}
and hence 
\begin{equation}
 \frac{\partial F}{\partial\ln\nu_c} 
       =  \frac{\partial F_0}{\partial\ln\nu_c}\,
          \left(1 + \frac{\sigma S_3}{6}H_3(\nu_c)
     - (\nu_c - \nu_c^{-1})\,\frac{\partial (\sigma S_3/6)}{\partial\ln\nu_c}
           \right)
       \approx  \frac{\partial F_0}{\partial\ln\nu_c}\,
          \left(1 + \frac{\sigma S_3}{6}H_3(\nu_c) \right)
 \label{eqn:vfvratio}
\end{equation}
\citep{fnlverde};  the term in brackets is the ratio of the halo mass
function when  $f_{nl}\ne 0$ to that when $f_{nl}=0$.   \cite[][argue
that, formally, the term in brackets is not the  full story, but that
it is, nevertheless, a good approximation.]{lsfnlhalo}  
The ratio
$(\partial F/\partial\ln\nu_c)/(\partial F_0/\partial\ln\nu_c)$ provides
a good description of the fractional change in the halo  mass function
induced by the coupling parameter $f_{nl}$ \citep{fnlvincent}, even
though the functional form of  $\partial F_0/\partial\ln\nu_c$ does
not provide a good description  of halo abundances in Gaussian
cosmologies.

Halo abundances in $f_{nl}=0$ simulations are usually well-approximated 
by the functional form of \cite{st99}:
\begin{equation}
\nu f(\nu) \equiv \frac{\partial F_0^{ST}}{\partial \ln \nu}
             = 2A\sqrt{a}\,\nu\, \left[1 + (\sqrt{a}\,\nu)^{-2p}\right] 
            \frac{e^{-a\nu^2/2}}{\sqrt{2\pi}}, \label{eqn:ST99}
\end{equation}
where $p=0.3$, $a=0.7$ and $A = 0.322$ comes from requiring that 
the integral over all $\nu$ equal unity.  Equation~(\ref{eqn:PS}) 
with $f_{nl}=0$ would yield $a=1$ and $p=0$, so $A$ would be 
modified appropriately.  
Recently, \citet{grossinfm} have studied the effect on the $f_{nl}$ 
model of simply setting $\delta_c\to\sqrt{a}\delta_c$ in 
equation~(\ref{eqn:PS}).
In effect, they ignore the consequences of $p\ne 0$.  
However, \cite{lsfnlhalo} have shown that 
when the non-Gaussianity is weak, then the correction factor 
for halo abundances
is well approximated by 
\begin{equation}
 \frac{\partial F/\partial \ln \nu_c}{\partial F_0/\partial \ln \nu_c}
  \approx 1 + \frac{\sigma S_3}{6}H_3\left(\frac{B(\sigma)}{\sigma}\right),
  \label{eqn:kcor}
\end{equation} 
where 
\begin{equation}
B(\sigma,z) = \sqrt{a}\,\delta_{\rm sc}(z)\, \left [1 + %
       \beta (\sqrt{a}\,\delta_{\rm sc}/\sigma)^{-2\alpha}\right],
       \label{eqn:STBS}
\end{equation}
with $a = 0.7$, $\beta = 0.485$, and $\alpha = 0.615$, 
is motivated by models of triaxial collapse \citep{smt01,st02}.

\begin{figure}
\centering
\includegraphics[width=0.8\linewidth]{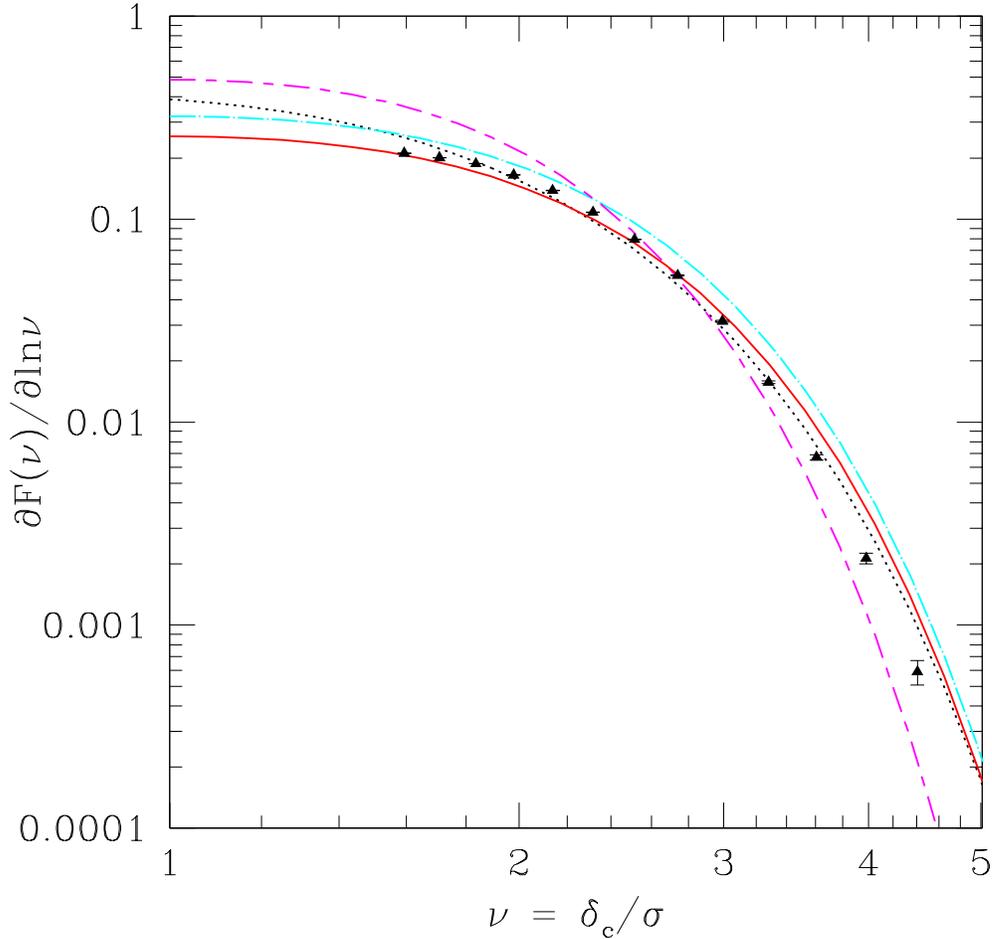}
\caption{Halo multiplicity function $\partial F/ \partial \ln \nu$ for the 
Gaussian simulations at $z =0.3$.  Black symbols show the measurements 
from simulations; short-long-dashed, dotted, dot-dashed and solid curves show 
equations~(\ref{eqn:PS}), (\ref{eqn:LS}), (\ref{eqn:ST99}), and 
(\ref{eqn:ST99}) with a new normalization ($A = 0.26$) respectively.}
\label{fig:MFgau}
\end{figure}

\subsection{If halo formation depends on all three initial eigenvalues exceeding a critical value}\label{all3}

Suppose that the criterion for halo formation is not that $\delta$, 
the sum of the eigenvalues, exceeds $\delta_c$, but that the smallest 
eigenvalue $\lambda_3$ exceeds $\lambda_c$.
Then, the analogous argument yields 
\begin{equation}
 F(>M) = F(<\sigma(R)) = \int_{3\lambda_c}^\infty 
          {\rm d}\delta\, p(\delta|R)\,
          \int_{\lambda_c}^{\delta/3} {\rm d}\lambda_3\, p(\lambda_{3}|\delta)
       = \int_{3\lambda_c}^\infty 
          {\rm d}\delta\, \left[1 + \frac{\sigma S_3}{6}H_3\right]\,
                    p_0(\delta|R)\,P_0(\lambda_{3}>\lambda_{c}|\delta),
\end{equation}
where 
\begin{equation}
 P_0(\lambda_{3}\ge\lambda_{c}|\delta)
 = \Bigg{\{}-\frac{3\sqrt{10}}{4\sqrt{\pi}}
      \frac{(\delta-3\lambda_{c})}{\sigma}
  \exp\bigg{[}-\frac{5(\delta-3\lambda_{c})^2}{8\sigma^2}\bigg{]}
  + \frac{1}{2}\bigg{\{}{\rm erf}\bigg{[}
      \frac{(\delta-3\lambda_{c})\sqrt{10}}{4\sigma}\bigg{]}
    + {\rm erf}\bigg{[}\frac{(\delta-3\lambda_{c})\sqrt{10}}
      {2\sigma}\bigg{]}\bigg{\}}\Bigg{\}}\Theta(\delta-3\lambda_{c}).
\end{equation}
If we define $\ell_c \equiv \lambda_c/\sigma(R)$, then 
\begin{eqnarray}
 F(\ge \ell_c) &\equiv& 
  F_0(\ge\ell_c)  + \frac{\sigma S_3}{6}\,\Delta F(\ge\ell_c), 
 \qquad {\rm where} \label{eqn:f0df}\\
 \Delta F(\ge\ell_c) &\equiv& \int_{3\ell_c}^\infty 
          {\rm d}\nu\,\frac{\exp(-\nu^2/2)}{\sqrt{2\pi}} \,\nu(\nu^2-3)\,
                    \,P_0(\nu - 3\ell_{c}) \nonumber\\
 &=& \frac{5^{3/2}}{162\sqrt{2\pi}}\,(100\ell_c^4 - 105\ell_c^2 + 9)
            \,\exp(-5\ell_c^2/2)\,{\rm erfc}(\sqrt{2}\ell_c)
     - \frac{5^{3/2}}{48\sqrt{3\pi}}\,(2 - 15\ell_c^2)
            \,\exp(-15\ell_c^2/4)\,{\rm erfc}(\sqrt{3}\ell_c/2)\nonumber\\
 && \qquad\qquad + \frac{125\sqrt{5}}{648\pi}\, \ell_c\,\exp(-9\ell_c^2/2)\, 
    \left(5 - 8\,\ell_c^2\right)
 \label{eqn:dflc}.
\end{eqnarray}
The halo mass function is 
\begin{eqnarray}
  \frac{\partial F}{\partial\ell_c} &=& \frac{\partial F_0}{\partial\ell_c} + 
  \frac{\partial\, (\sigma S_3/6)}{\partial\ell_c}\, \Delta F(\ge\ell_c) + 
  \frac{\sigma S_3}{6} \frac{\partial\, (\Delta F)}{\partial\ell_c}\nonumber\\
 \frac{\partial F_0}{\partial\ell_c} &=& 
  -\frac{\sqrt{10}}{\sqrt{\pi}}\, 
   \left(\frac{5\ell_c^2}{3} - \frac{1}{12}\right)\,
   \exp\left(-\frac{5\ell_c^2}{2}\right)\,{\rm erfc}(\sqrt{2}\ell_c) - 
  \frac{\sqrt{15}}{4\sqrt{\pi}}\, \exp\left(-\frac{15\ell_c^2}{4}\right)\,
   {\rm erfc}\left(\frac{\sqrt{3}\ell_c}{2}\right) + 
  \frac{5\sqrt{5}}{3\pi}\,\ell_c\,\exp\left(-\frac{9\ell_c^2}{2}\right)
  \nonumber\\
 \frac{\partial\, (\Delta F)}{\partial\ell_c} &=& 
 \frac{25}{2}\frac{\sqrt{5}}{2^4 3^4\pi}\Biggl\{
   \exp\left(-\frac{9\ell_c^2}{2}\right)\, (64 - 870\ell_c^2 + 800\ell_c^4) - 
    \sqrt{2\pi}\, 8\,\ell_c(51 - 185\ell_c^2 + 100\ell_c^4)
    \,\exp\left(-\frac{5\ell_c^2}{2}\right)
    \,{\rm erfc}(\sqrt{2}\ell_c) + \nonumber\\
 && \qquad\qquad \sqrt{3\pi}\,3^4\ell_c\,(2 - 5\ell_c^2)\,
   \exp\left(-\frac{15\ell_c^2}{4}\right)\,
     {\rm erfc}\left(\frac{\sqrt{3}\ell_c}{2}\right) \Biggr\}.
 \label{eqn:LS}
\end{eqnarray}
Notice that 
\begin{equation}
 \frac{\partial F}{\partial\ell_c} \ne 
 \frac{\partial F_0}{\partial\ell_c}\,
  \left(1 + \frac{\sigma S_3}{6}\,H_3(\ell_c) 
        - \left(\ell_c-\ell_c^{-1}\right) 
          \frac{\partial\, (\sigma S_3/6)}{\partial\ln\ell_c}\right).
 \label{eqn:lflratio}
\end{equation}
Thus, in this case, the $f_{nl}$ modification to the halo mass function 
is qualitatively different from that associated with the spherical 
evolution model.  

We mentioned above that the ratio 
 $(\partial F/\partial \ln\nu_c)/(\partial F_0/\partial \ln\nu_c)$ is 
well described by the term in brackets in equation~(\ref{eqn:vfvratio}).
Since $\nu_c$ and $\ell_c$ are linearly proportional to one-another, 
it is interesting to ask if 
 $(\partial F/\partial \ln\ell_c)/(\partial F_0/\partial \ln\ell_c)$ 
is also well described by the term in brackets in 
equation~(\ref{eqn:vfvratio}).  
(We have already shown that this ratio is not described by simply replacing 
$\nu_c \to 3\ell_c$.)  Figure~\ref{fig:MFratio} shows that it is not.  
Thus, if we were certain that the logic which leads to this estimate 
of the mass function were reliable, then we would conclude that, by 
studying how halo abundances depend on $f_{nl}$, we may have learnt 
something important about the physics of halo formation:  the initial 
overdensity matters more than the value of the smallest eigenvalue.


\subsection{Comparison with measurements from numerical simulations}
Figure~\ref{fig:MFgau} shows the multiplicity function measured in 
the $f_{nl}=0$ simulations of \cite{fnlvincent}, where a detailed 
description of the runs can be found.  
(Our analysis is complementary to that of \citep{pphfnl08}, who 
have recently run a large set of simulations of the $f_{nl}$ model; 
they studied halo abundances {\em and} clustering in their simulations.)  
Curves show equation~(\ref{eqn:PS}) (magenta, short-long-dashed), 
equation~(\ref{eqn:LS}) (black, dotted),
and equation~(\ref{eqn:ST99}) (cyan, dashed)
with $\delta_c = 1.66$ and $\lambda_c = 0.41$.
In contrast to most previous work, equation~(\ref{eqn:LS}) appears to 
give a better fit than equation~(\ref{eqn:ST99}); this may be due to 
the fact that the halo finder ({\small AHF}, see \cite{kk09}) is not 
standard.  (The halo-finder used by \cite{pphfnl08} is more standard, 
and they indeed find that equation~\ref{eqn:ST99}, with standard choices 
for its free parameters, works well.)
The red solid curve shows equation~\eqref{eqn:ST99} with a new 
normalization ($A = 0.26$) and the agreement with the numerical 
measurements is much better.

\begin{figure}
\centering
 \includegraphics[width=0.47\linewidth]{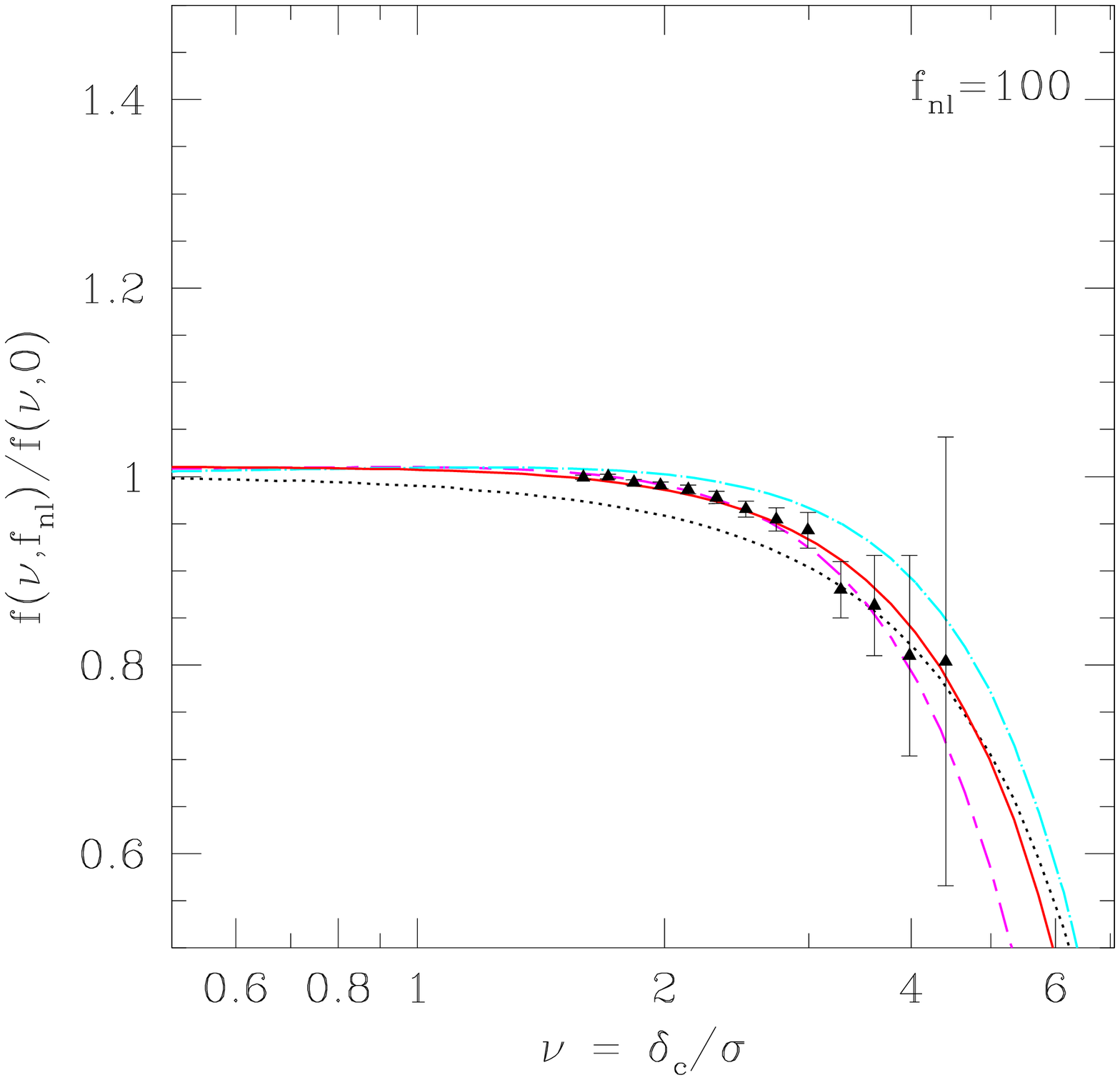}
 \includegraphics[width=0.47\linewidth]{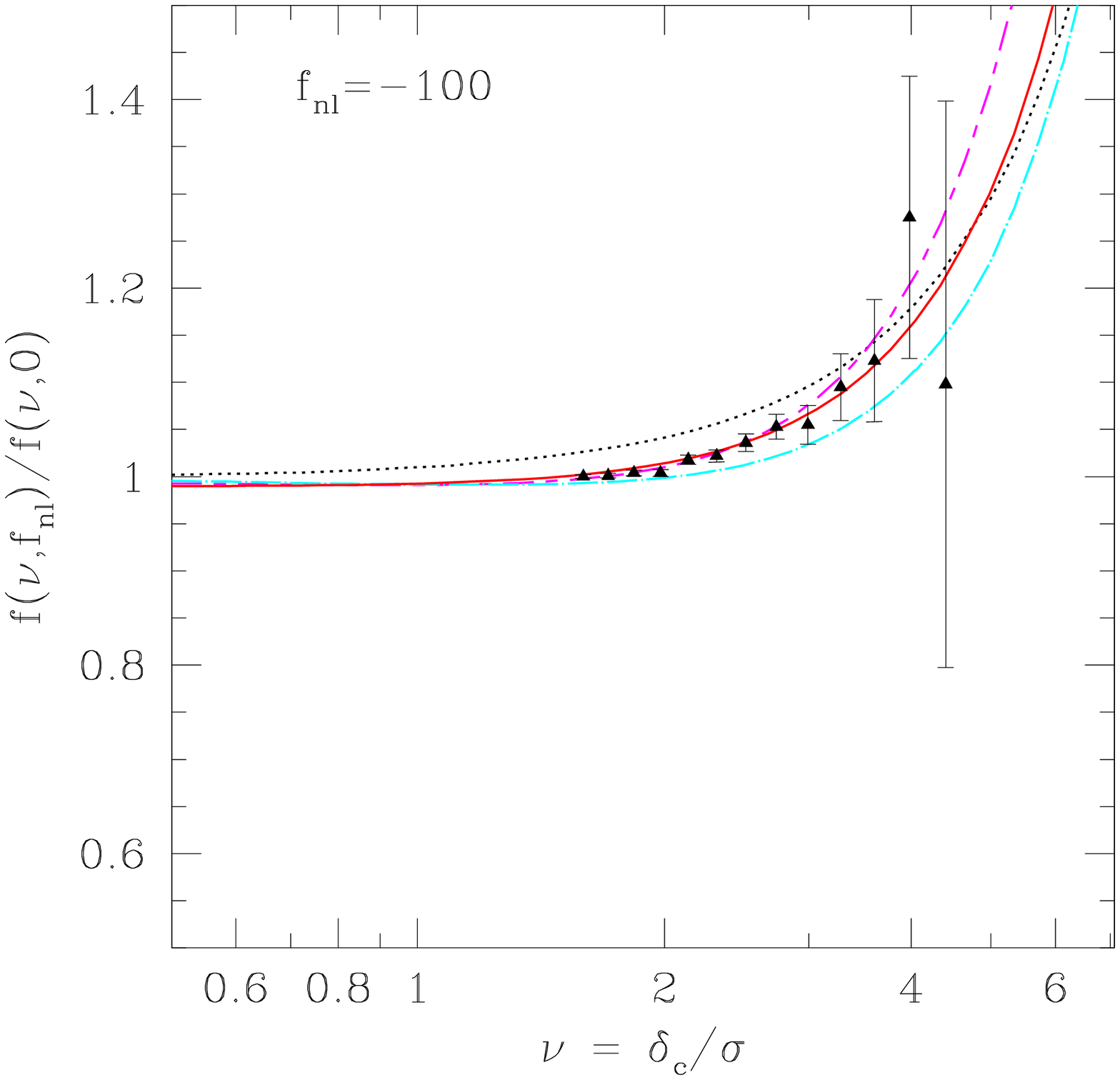}
 \caption{The ratio of halo mass function of $f_{nl}=100$ models to the 
   corresponding Gaussian models using equations~(\ref{eqn:vfvratio}), 
   (\ref{eqn:kcor}), \eqref{eqn:STPSra} and~(\ref{eqn:LS}), 
   (magenta short-long-dashed, solid red, cyan dot-dashed, and black dotted). 
   Panel on left shows $f_{nl}=100$ and panel on right shows $f_{nl}=-100$.}
 \label{fig:MFratio}
\end{figure}

It is conventional to show the effects of $f_{nl}\ne 0$ on the mass 
function by ratioing with respect to the $f_{nl}=0$ case.  
The symbols in Figures~\ref{fig:MFratio} show this ratio for 
$f_{nl}=100$ (left) and $-100$ (right) respectively.  
The short-long-dashed, dotted and dot-dashed curves show these 
ratios for the same models as in the previous figure 
(equations~\ref{eqn:vfvratio}, \ref{eqn:LS} and~\ref{eqn:STPSra}).  
The solid (red) curve shows equation~(\ref{eqn:kcor}).
The estimate motivated by the spherical collapse model 
(equations~\ref{eqn:vfvratio}) describes the measured ratio very well 
\citep[in agreement with][]{fnlverde,fnlvincent,grossinfm}, even though 
the mass function on which it is based is a bad fit to the $f_{nl}=0$ data.  
On the other hand, the same logic applied to a prescription based on the 
smallest eigenvalue (equation~\ref{eqn:LS}) fits the Gaussian 
(i.e. $f_{nl}=0$) mass function reasonably well, but does not describe 
deviations from non-Gaussianity very well!
In addition, the same logic applied to the \cite{st99} mass function 
(equation~\ref{eqn:STPSra}) also does not fit the ratio very well.  
However, our excursion set based approach (equation~\ref{eqn:kcor}) 
seems to match the measurement as well as, if not better than, any 
of the other methods.  
Note that it is the only model which matches 
both the $f_{nl}=0$ mass function, {\em and} the $f_{nl}\ne 0$ ratio.
We also checked that our excursion set based approach agrees with \citet{pphfnl08} 
fitting formula's prediction  (less than 6\% difference in the range of validity 
of the fitting formula, which is $1.4 < \nu < 5$).

\section{Void abundances}\label{section:vf}
Underdense regions are also a good probe of $f_{nl}$ \citep{lamshethfnl}.  
\citet{kvj08} have applied the analog of equation~(\ref{eqn:PS}) to 
study void abundances when $f_{nl}\ne 0$.  
In what follows, we estimate void abundances associated with the 
analog of Section~\ref{all3} by assuming they are simply the opposite 
of halos.  However, because of the `void-in-cloud' problem identified 
by \cite{sw04}, even when $f_{nl}=0$, these analyses are, at best, 
appropriate only for the largest voids.   
Hence, we also discuss the effect of including the correction for the 
`void-in-cloud' problem.  

\subsection{If voids formation depends on all three eigenvalues lying 
 below a critical value}
Given a critical value $\lambda_v$, the corresponding mass function 
of voids is 
\begin{equation}
F(< M) = \int^{3\lambda_v}_{-\infty} d\delta\, p(\delta|R)\, 
         \int^{\lambda_v}_{\delta/3} d\lambda_1\, p(\lambda_1|\delta)
       = \int_{-\infty}^{3\lambda_v}d\delta\, 
            \left[1 + \frac{\sigma S_3}{6}H_3\right]
           p_0(\delta|R)P_0(\lambda_1<\lambda_v|\delta),
\end{equation}
where
\begin{equation}
P_0(\lambda_1 \le \lambda_v|\delta) = \left\{-\frac{3\sqrt{10}}{4\sqrt{\pi}} %
   \frac{(3\lambda_v-\delta)}{\sigma}  %
   \exp\left[-\frac{5(3\lambda_v-\delta)^2}{8\sigma^2}\right] + %
    \frac{1}{2}\left\{ %
       {\rm erf}\left[\frac{(3\lambda_v-\delta)\sqrt{10}}{4\sigma}\right] %
       {\rm erf}\left[\frac{(3\lambda_v-\delta)\sqrt{10}}{2\sigma}\right] %
\right\}
\right\}\Theta(3\lambda_v-\delta).
\end{equation}
If we define $\ell_v\equiv \lambda_v/\sigma$, then the corresponding 
$F_0(\le \ell_v)$ and $\Delta F(\le \ell_v)$ can be obtained by replacing 
${\rm erfc}(x)$ by $-[1 + {\rm erf}(x)]$ in equations~(\ref{eqn:f0df}) and 
(\ref{eqn:dflc}). 
The void mass function can be obtained by making the same replacement, 
so it is 
\begin{align}
\frac{\partial F}{\partial \ell_v}  = & \frac{\partial F_0}{\partial \ell_v}
     + \frac{\partial (\sigma S_3/6)}{\partial \ell_v}\Delta F(\le \ell_v)
     + \frac{\sigma S_3}{6}\frac{\partial (\Delta F)}{\partial \ell_v} \nonumber\\
\frac{\partial F_0}{\partial \ell_v}  = & \frac{\sqrt{10}}{\sqrt{\pi}}
   \left(\frac{5\ell_v^2}{3}-\frac{1}{12}\right)\exp\left(-\frac{5\ell_v^2}{2}\right)
   \left[1 + {\rm erf}(\sqrt{2}\ell_v)\right] +
   \frac{\sqrt{15}}{4\sqrt{\pi}}\exp\left(-\frac{15\ell_v^2}{4}\right)
   \left[1 + {\rm erf}\left(\frac{\sqrt{3}\ell_v}{2}\right)\right] +
   \frac{5\sqrt{5}}{3\pi}\ell_v\exp\left(-\frac{9\ell_v^2}{2}\right) \nonumber\\
\frac{\partial (\Delta F)}{\partial \ell_v}  = &\frac{25}{2}
   \frac{\sqrt{5}}{2^43^4\pi}\left\{\exp\left(-\frac{9\ell_v^2}{2}\right)  %
   (64-870\ell_v^2+800\ell_v^4) +\sqrt{2\pi}8\ell_v(51-185\ell_v^2+100\ell_v^4)        %
   \exp\left(-\frac{5\ell_v^2}{2}\right)\left[1 + {\rm erf}(\sqrt{2}\ell_v)\right] %
   \right. \nonumber \\
 & \left. - \sqrt{3\pi}3^4\ell_v(2-5\ell_v^2)\exp\left(-\frac{15\ell_v^2}{4}\right)  %
   \left[1 + {\rm erf}\left(\frac{\sqrt{3}\ell_v}{2}\right)\right]         %
\right\}.
 \label{eqn:LSvoids}
\end{align}

\subsection{Void-in-cloud problem: excursion set approach}
An important aspect in the void abundance is the overcounting of the voids 
located inside collapsing regions. The formalism of counting voids 
as regions below some critical value 
(denoted $\delta_v$ in \citet{kvj08} and $\lambda_v$ in the discussion above)
does not account for this. 
\citet{sw04} examined this problem using the excursion set approach by 
studying a two barriers problem: $\delta_c$ for haloes and $\delta_v$ for 
voids.

We will now extend the calculation of the void-in-cloud effect to 
models where $f_{nl} \neq 0$. 
We use the constant barriers to demonstrate the method.
Denote the two constant barriers correspond to the 
formation of halos and voids by $\delta_c$ and $\delta_v$,
$\mathcal{F}(s,\delta_v,\delta_c)$ as the probability of a random walk 
crossing the barrier $\delta_v$ at scale $s$ and it did not cross the other 
barrier $\delta_c$. 
This probability is directly connected to the void abundances, including the 
void-in-cloud effect, as the random walk never crossed the halo formation 
barrier. It is related to the first crossing distribution $f(s,\delta_v)$ by
\begin{equation}
\mathcal{F}(s,\delta_v,\delta_c) = f(s,\delta_v) - 
    \int_0^s{\rm d}S_1\, \mathcal{F}(S_1,\delta_c,\delta_v)
                          f(s,\delta_v | S_1,\delta_c),
\label{eqn:fsdvdc}
\end{equation}
where the second term on the right hand side substracts from the first 
crossing those walks that  crossed $\delta_c$ at $S_1$ 
before crossing $\delta_v$ at $s$ 
(but never had crossed $\delta_v$ before $S_1$).
Swaping $\delta_v$ and $\delta_c$:
\begin{equation}
\mathcal{F}(S_1,\delta_c,\delta_v) = f(S_1,\delta_c) - 
    \int_0^{S_1}{\rm d}S_2\, \mathcal{F}(S_2,\delta_v,\delta_c)
                          f(S_1,\delta_c | S_2,\delta_v).
\label{eqn:fsdcdv}
\end{equation}
Substituting equation~\eqref{eqn:fsdcdv} into equation~\eqref{eqn:fsdvdc},
we find
\begin{align}
\mathcal{F}(s,\delta_v,\delta_c) & = f(s,\delta_v) - 
 \int_0^s {\rm d}S_1\, f(s,\delta_v | S_1,\delta_c)f(S_1,\delta_c)
 +  \int_0^s{\rm d}S_1 \int_0^{S_1}{\rm d}S_2
    f(s,\delta_v | S_1,\delta_c)f(S_1,\delta_c | S_2,\delta_v)
\mathcal{F}(S_2,\delta_v,\delta_c) \\
& = f(s,\delta_v) + \sum_{n=1}^{\infty} (-1)^n \int_0^{S_0}{\rm d}S_1 \dots 
\int_0^{S_{n-1}}{\rm d}S_n 
   \prod_{m=0}^{n-1} f(S_m,\delta_m|S_{m+1},\delta_{m+1}) f(S_n,\delta_n),
\label{eqn:FSint}
\end{align}
where the last expression is obtained after inserting
equations~\eqref{eqn:fsdvdc} and \eqref{eqn:fsdcdv} successively.  
Furthermore, $S_0 \equiv s$ and 
\begin{equation}
\delta_n = \left\{\begin{array}{rl}
 \delta_v & \text{if } n \text{ is even} \\
 \delta_c & \text{if } n \text{ is odd} 
\end{array} \right. 
\end{equation}
The $n$th order term in the summation of equation~\eqref{eqn:FSint} 
denotes  walks that have crossed the two barriers alternatively $n$ times 
before crossing $\delta_v$ at $s$.
Below we will work out the predictions of equation~\eqref{eqn:FSint}
for primordial Gaussian  perturbations and for models with 
primordial non-Gaussianity of the $f_{nl}$ type.

\subsubsection{Gaussian initial conditions}
We would like to estimate $f(s,\delta_v|S,\delta_c)$, which is 
the first crossing probability of $\delta_v$ at scale $s$ given 
it crossed the barrier $\delta_c$ at some scale $S (<s)$.
For Gaussian distributions with sharp-$k$ space filters, 
$f_0(s,\delta_v | S , \delta_c) = f_0(s -S, \delta_v - \delta_c)$,
and equation~\eqref{eqn:FSint} reduces to
\begin{align}
\mathcal{F}_0(s,\delta_v,\delta_c) & = 
 f_0(s,\delta_v) + \sum_{n=1}^{\infty} (-1)^n \int_0^{S_0}{\rm d}S_1 \dots 
\int_0^{S_{n-1}}{\rm d}S_n 
   \left[\prod_{m=0}^{n-1} f_0(S_m,\delta_m|S_{m+1},\delta_{m+1}) \right]
    f_0(S_n,\delta_n),
                                                  \nonumber \\
 & \approx f_0(s,\delta_v)
     \exp\left(-\frac{|\delta_v|}{\delta_c}\frac{\mathcal{D}^2}{4\nu^2} 
   - 2\frac{\mathcal{D}^4}{\nu^4}\right),
\label{eqn:swvoid}
\end{align}
where the last expression is the approximation given by \citet{sw04}
with $\mathcal{D}\equiv |\delta_v|/(\delta_c + |\delta_v|)$ and 
$\nu \equiv \delta_v/\sqrt{s}$.

\subsubsection{Local non-Gaussian $f_{nl}$ models}

The calculation of the conditional first crossing probability 
$f(\delta_v,s|\delta_c,S)$ for the $f_{nl}$ model is analogous to that of
halo abundances \citep{lsfnlhalo}.
First we write down the probability $p(s,\delta | S,\delta_v)$ as
\begin{equation}
p(s,\delta |S,\delta_c) = \int^s_S {\rm d}S'\, 
                f(S',\delta_v|S,\delta_c)\,p(s,\delta|S',\delta_v|S,\delta_c),
\end{equation}
provided that $\delta < \delta_v$. So,
\begin{equation}
P(s,\delta_v|S,\delta_c) 
   \equiv \int^{\delta_v}_{-\infty}{\rm d}\delta\, p(s,\delta |S,\delta_c) =
       \int^s_S {\rm d}S'\, f(S',\delta_v|S,\delta_c)\,
    \int^{\delta_v}_{-\infty}{\rm d}\delta\,p(s,\delta|S',\delta_v|S,\delta_c).
\end{equation}
The derivative with respect to $s$ is
\begin{equation}
\frac{\partial P(s,\delta_v|S,\delta_c)}{\partial s} = 
  \frac{f(s,\delta_v|S,\delta_c)}{2} +  
    \int^s_S {\rm d}S'\, f(S',\delta_v|S,\delta_c)\,\frac{\partial}{\partial s}
   \int^{\delta_v}_{-\infty}{\rm d}\delta\,p(s,\delta|S',\delta_v|S,\delta_c).
\end{equation}
The above equation is an integral equation for $f(\delta_v,s|\delta_c,S)$ and
its zeroth-order solution is given by the left hand side of the equation 
(which can be evaluated using the bivariate Edgeworth expansion).
We argue that, in analogy to the calculation of the halo abundance, 
the first-order solution is negligible compared to the zeroth-order. 
Therefore, we can make the following approximation
\begin{equation}
f(s,\delta_v|S,\delta_c) \approx 
 2\frac{\partial P(s,\delta_v|S,\delta_c)}{\partial s}.
\label{eqn:con1stcrossing}
\end{equation}
Note that for Gaussian distributions,
\begin{equation}
\begin{split}
\frac{\partial P_0(s,\delta_v|S,\delta_c)}{\partial s}
& = \frac{\partial}{\partial s}\int^{(\delta_v-\delta_c)/\sqrt{s-S}}_{-\infty} 
     {\rm d}x\, \frac{e^{-x^2/2}}{\sqrt{2\pi}} \\
& = \frac{-(\delta_v-\delta_c)}{2(s-S)}
       \frac{\exp[-(\delta_v-\delta_c)^2/2(s-S)]}{\sqrt{2\pi(s-S)}},
\end{split}
\end{equation}
which is the expected conditional distribution for Gaussian statistics.

In Appendix~\ref{section:beeva}, the right hand side of 
equation~\eqref{eqn:con1stcrossing} is evaluated
using the Edgeworth expansion. Here, we will approximate 
the conditional first crossing probability by 
\begin{equation}
f(s,\delta_v|S,\delta_c) \approx f_0(s,\delta_v|S,\delta_c)
\left[1 + \frac{\sigma S_3}{6}\,\zeta(s,\delta_v,S,\delta_c)\right],
\label{eqn:con1st}
\end{equation}
where 
\begin{equation}
\zeta(s,\delta_v,S,\delta_c) = -2 \frac{\partial \mathcal{E}(s,S)}{\partial s}
\frac{(s-S)^{3/2}}{|\delta_v - \delta_c|} 
      - \mathcal{E}(s,S)\frac{|\delta_v -\delta_c|}{\sqrt{s-S}}.
\end{equation}
Substituting equation~\eqref{eqn:con1st} into equation~\eqref{eqn:FSint} and
keeping only terms linear in $(\sigma S_3/6)$, we can recast the integral
equation into the form
\begin{equation}
\begin{split}
&\mathcal{F}(s,\delta_v,\delta_c) =  \mathcal{F}_0(s,\delta_v,\delta_c) 
+ \frac{\sigma S_3}{6}
  \Bigg\{ f_0(s,\delta_v) H_3\left(\frac{\delta_v}{\sqrt{s}}\right)
                                                       \\
&  + \sum_{n=1}^{\infty}(-1)^n \int_0^{S_0}{\rm d}S_1 \dots
   \int_0^{S_{n-1}}{\rm d}S_n 
   \left[\prod_{m=0}^{n-1} f_0(S_m,\delta_m | S_{m+1},\delta_{m+1})\right]
   f_0(S_n,\delta_n)\left[ 
\sum_{m=0}^{n-1}\zeta(S_m,\delta_m,S_{m+1},\delta_{m+1}) + 
   H_3\left(\frac{\delta_n}{\sqrt{S_n}}\right)\right]
  \Bigg\}.
\end{split}
\end{equation}

\subsection{Comparison of models}
\begin{figure}
\centering
\includegraphics[width=0.8\textwidth]{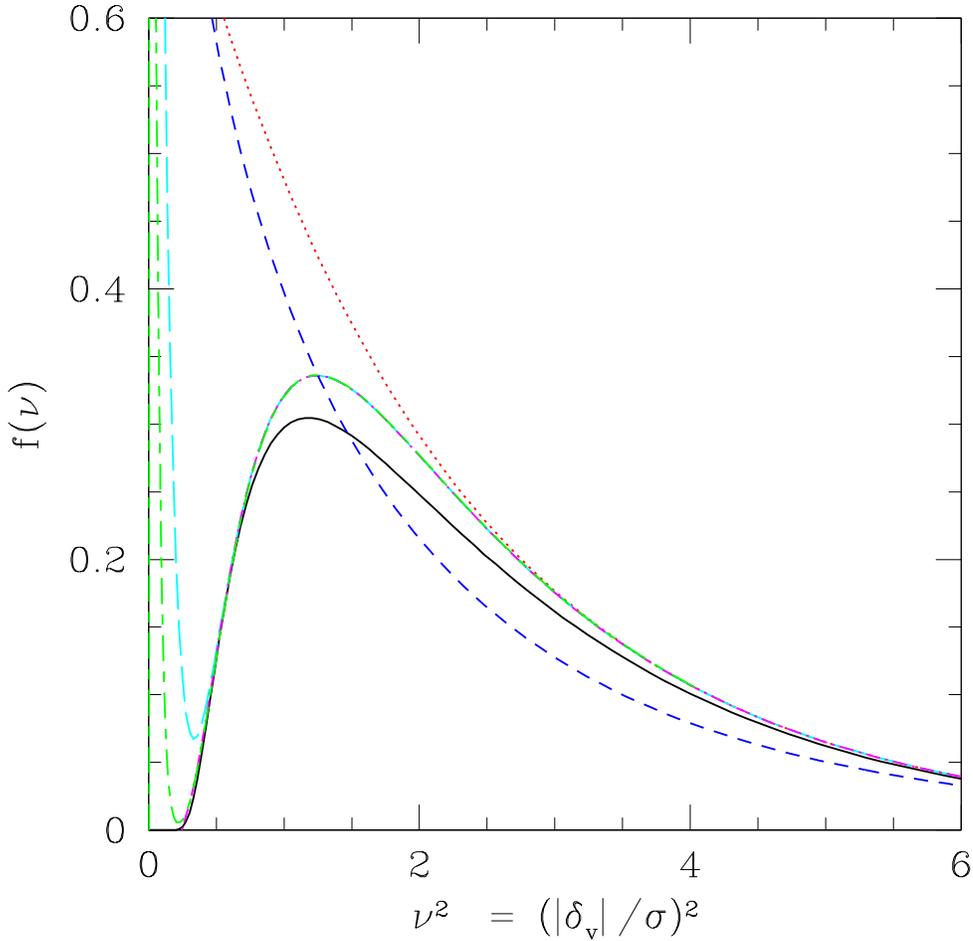}
\caption{Predicted void abundances for Gaussian initial conditions.
        Solid (black) curve shows equation~\eqref{eqn:swvoid} 
        with $(\delta_v,\delta_c) = (-2.81,1.66)$. 
        Long-dashed (cyan), dot-dashed (magenta), and short-long-dashed (green)
        curves are the solution to equation~\eqref{eqn:FSint} with 
        $n =2, 3, 4$ respectively. 
        Predictions without the void-in-cloud effect are also shown: 
        dotted (red) is the zeroth order solution of
        equation~\eqref{eqn:FSint}, which is equivalent to the PS result; 
        short-dashed (blue) shows the prediction from the LS formalism 
        (equation~(\ref{eqn:LSvoids}) with $\lambda_v = -0.69$).
        }
\label{fig:fnu}
\end{figure}
Figure~\ref{fig:fnu} shows theoretical expectations of void abundances when $f_{nl}=0$:
The solid (black) curve shows equation~\eqref{eqn:swvoid} 
with $(\delta_v,\delta_c) = (-2.81,1.66)$; The
Long-dashed (cyan), dot-dashed (magenta), and short-long-dashed (green) curves
show equation~\eqref{eqn:FSint}, keeping terms up to 
$n =2, 3, 4$ respectively.
The zeroth order solution of equation~\eqref{eqn:FSint} without the void-in-cloud effect
is shown as the dotted (red) curve. It is the first crossing probability of a constant barrier 
and is the same as the PS prediction.
Finally, the short-dashed (blue) shows the prediction of the LS formalism 
(equation~\ref{eqn:LSvoids}) with $\lambda_v = -0.69$.

The predictions from the PS formalism and the LS formalism (both
without the  void-in-cloud effect) are different over  a large range
of void size, so comparisons with numerical measurements  could
distinguish which models describe best the formation of voids.
Furthermore, while the effect of void-in-cloud is significant 
for small voids, it is negligible for big voids.

The approximation equation~\eqref{eqn:swvoid} provides a very good
description of the solution to the integral
equation~\eqref{eqn:FSint}, even in the regime of very small voids
($\nu^2 \sim 0.5)$ for which equation~\eqref{eqn:FSint} requires  the
inclusion of high order terms.  If  one characterises the accuracy of the
$n$-order term by the smallest $\nu$  at which the inclusion of the
next order term modifies the result by less than  $1\%$, then the
second and third orders are accurate for $\nu \gtrsim 0.9$ and  $\nu
\gtrsim 0.6$.

\begin{figure}
 \centering
 \includegraphics[width=0.47\linewidth]{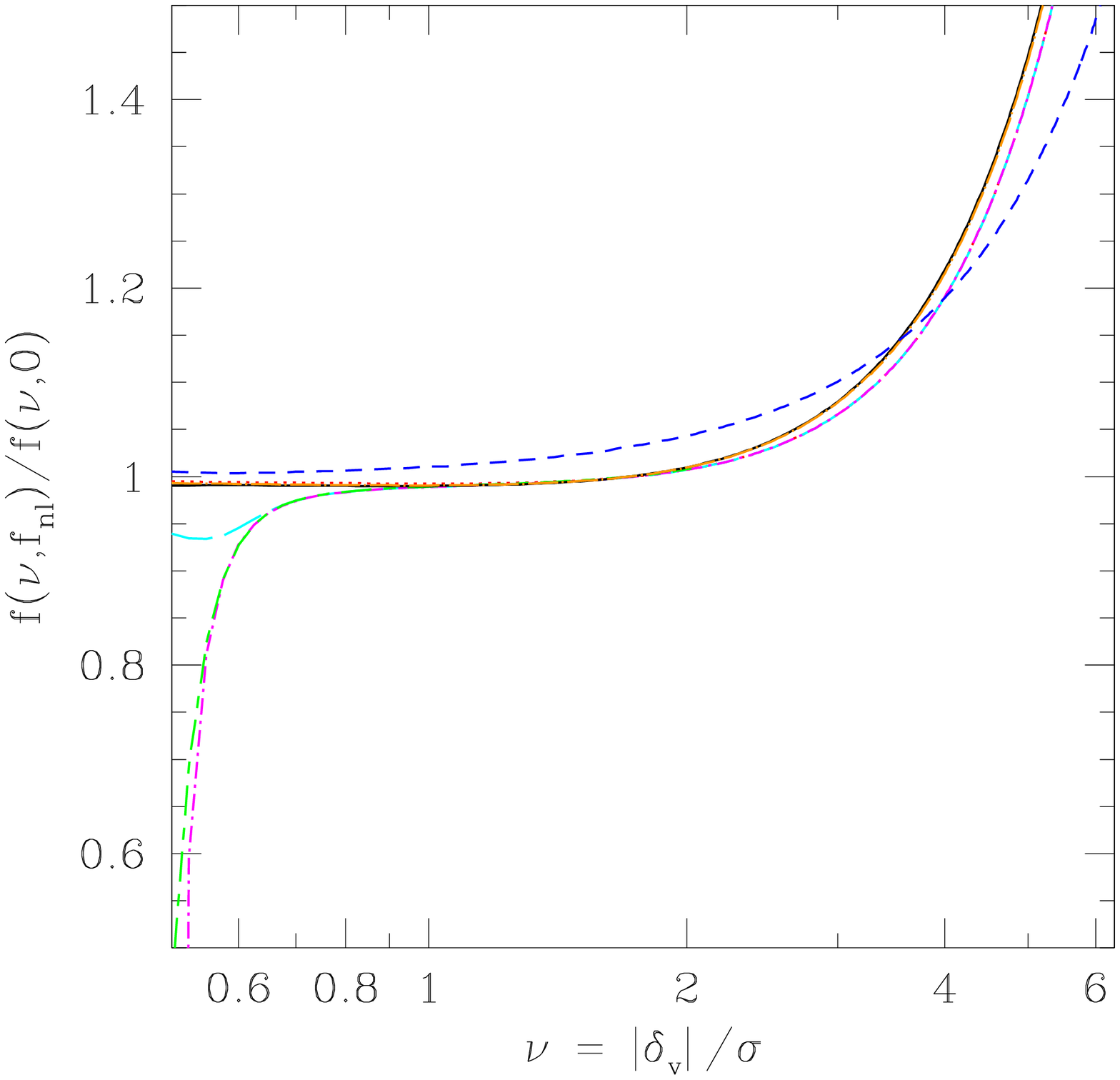}
 \includegraphics[width=0.47\linewidth]{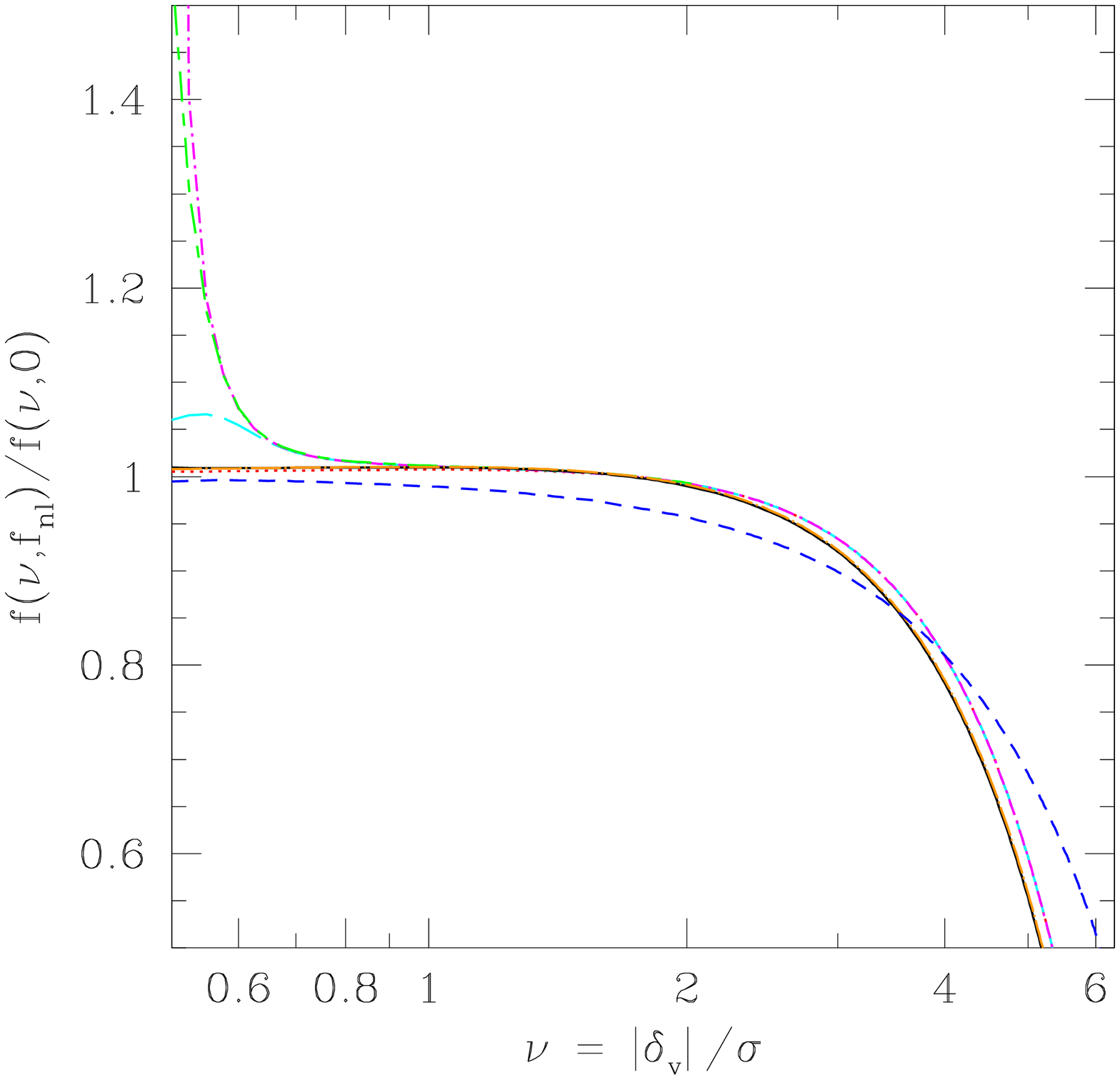}
 \caption{Ratio of void abundances of $f_{nl}\neq 0$ to the $f_{nl}= 0$ case. 
         Left and right panels show $f_{nl}=100$ and $-100$ respectively.
         Curve labels are the same as the previous figure, with 
         the exception of the solid (black) showing the PS 
         prediction (the square brackets in equation~\eqref{eqn:vfvratio}) 
         and an addition 
         dot-long-dashed (orange) curve showing the same equation
         but ignoring the $\partial (\sigma S_3)/\partial s$ term.          
          Models without the void-in-cloud effect (black solid, 
           blue short-dashed, and orange dot-long-dashed) 
          include the scale dependence of $\sigma S_3$; the others 
          use $|\sigma S_3| = 0.022$.
          }
 \label{fig:VFratio}
\end{figure}
Figure~\ref{fig:VFratio} compares the ratio of the void abundances 
for $f_{nl}\neq 0$ relative to the case $f_{nl}=0$ for the various 
analytic approximations described above.
The curve labels are identical to those in the previous figures, 
except for the solid (black) curve showing equation~\eqref{eqn:vfvratio}
and an
addition dot-long-dashed (orange) curve showing the same equation
upon neglecting the $\partial (\sigma S_3)/\partial s$ term.
These two curves, as well as the short-dashed (blue) curve, include 
the scale dependence of $\sigma S_3$ using the approximation formula given 
in \citet{lsfnlhalo}.
The other curves (dotted (red), long-dashed (cyan), dot-short-dashed (magenta),
 and short-long-dashed (green) for $n=0,2,3,4$ in equation~\ref{eqn:FSint}) 
assume a constant $\sigma S_3=0.022$ 
As we can see, unlike halo abundances, a positive value of $f_{nl}$ 
increases the relative number of big voids, whereas a negative $f_{nl}$
decreases it.

The overlapping of the solid (black), dot-long-dashed (orange), and
dotted (red) curves justifies a posteriori our assumption of constant
$\sigma S_3$.  For big voids, the dotted curve deviates only slightly
from the other two where it agrees reasonably well with higher order
solulions to the integral  equation~\eqref{eqn:FSint}.  Hence, the
difference between the various curves (apart from the blue
short-dashed)  are mostly governed by $\sigma S_3$ rather than the
different functional form of the models. Note, however, that the
prediction of the LS formalism (blue short-dashed)  considerably
departs from the other curves.

Including higher order terms in equation~\eqref{eqn:FSint} so as to
better account for the void-in-cloud effect does not change the 
results for big voids. This is expected from our model in which 
big voids are unlikely to be embedded in a larger collapsing region.
However, including the void-in-cloud effect modifies the ratio of
void abundances for the smallest voids. Namely, a positive $f_{nl}$ tend
to decreases the number of small voids.
This may originate from an increase of high mass halos, which is such that 
small voids are more likely to sit inside an collapsing region.
Our prediction is at best qualitative because 
higher order terms are needed to describe accurately the small voids regime
(Our results shown in figure~\ref{fig:fnu} indicate that the third order 
approximation (magenta) is valid only for $\nu \gtrsim 0.6$). 
We have not pursued the inclusion of higher order corrections here since 
constant barriers may not be a good approximation for the formation of 
halos and voids as suggested by the analysis of (Gaussian) initial 
conditions of cosmological simulations.
Nonetheless, our formalism  allows the incorporation of 
scale dependence barriers, which shall be useful when numerical measurements 
are available.

\section{Discussion}
We extended Doroshkevich's celebrated formulae for the eigenvalues 
of the initial shear field associated with Gaussian statistics to 
the local non-Gaussian $f_{nl}$ model. 
We showed that, up to second order in $f_{nl}$, this is 
straightforward because, at fixed overdensity, the distribution 
is the same as when $f_{nl}=0$ (i.e., Gaussian initial conditions).  
Our analytic formulae are in good agreement with measurements 
of the distribution of $(\lambda_1,\lambda_2,\lambda_3)$ in 
Monte Carlo realizations of the $f_{nl}$ distribution
(Figures~\ref{fig:initpdfs} and~\ref{fig:initge}).  

Our extension of Doroshkevich's formulae to the local non-Gaussian 
$f_{nl}$ model provides the first step in the study of triaxial 
structure formation in models with primordial non-Gaussianity.
This is interesting because, for Gaussian initial conditions, halo 
formation is more triaxial than spherical.  
In particular, in the triaxial collapse model, the evolution of 
a patch depends on its initial overdensity $\delta$ as well as the 
parameters $e$ and $p$, which describe its initial ellipticity and 
prolateness.  We showed that the distribution of $e$ and $p$ at 
fixed $\delta$ is unchanged from when $f_{nl}=0$ (equation~\ref{gepd}).  
Therefore, equation~(\ref{eqn:STBS}), which was determined for 
$f_{nl}=0$ models, should continue to be useful even when $f_{nl}\ne 0$.  

We applied our formulae for the initial shear field to study the 
change in halo and void abundances in the local non-Gaussian model 
(Section~\ref{section:mf}).  
When $f_{nl}=0$, halo abundances predicted by a model in which halo formation 
is associated with having all three initial shear eigenvalues above 
some critical value are in better agreement with the simulations than
those implied by the usual overdensity threshold criterion 
(Figure~\ref{fig:MFgau}).  
However, the predicted dependence of $f(\nu,f_{nl})/f(\nu,0)$ $f_{nl}$ is
in better agreement with the simulations (see Fig.~\ref{fig:MFratio}) in
the latter case.

To understand this better, we extended the ellipsoidal collapse 
formalism to the local non-Gaussian model. 
We included the moving barrier formulation of ellipsoidal collapse 
\citep{smt01,st02} using two different approaches.  
The first is analogous to that used in the case of spherical 
collapse (Section~\ref{section:halocritical}); the predicted 
dependence on $f_{nl}$ did not agree with measurements from the 
simulations.
The second is an extension of the excursion set approach 
following \cite{lsfnlhalo}.  
For the case of a constant barrier (associated with spherical
collapse), this approach reproduces the results of \cite{fnlverde}
and \cite{mr09c}. Its extension to moving barriers appears promising
since it matches the measured halo counts when $f_{nl}=0$ as well as
the  dependence on $f_{nl}$ (Figure~\ref{fig:MFratio}).

For $f_{nl}$ models, differences in the density field evolved from
Gaussian and non-Gaussian initial conditions are more dramatic in the
underdense regions \citep{lamshethfnl}.  Figure~\ref{fig:fnu} shows a
number of predictions for void  abundances for $f_{nl}=0$, while
Figure~\ref{fig:VFratio} shows the  effect of using our triaxial
formalism to study how these depend on $f_{nl}$.  The trends are
generally like those in the halo abundance, except that the dependence
on the  sign of $f_{nl}$ is reversed (compare
Figures~\ref{fig:MFratio}  and~\ref{fig:VFratio}). This is consistent
with the recent work of \cite{kvj08}.   Still, one might have expected
that an excess of massive halos also implies an excess of large voids
(since all the mass is concentrated in a smaller  volume).  This is
what is indeed found for $f_{nl}=0$ when one increases the
normalisation amplitude  $\sigma_8$ of the fluctuation field. However,
this is  not true in $f_{nl}$ models.   In this respect, void
abundances provide complementary information  to cluster abundances,
so a joint comparison of both could be used  to put constraints on the
level of non-Gaussianity.
  
We also demonstrated how the void-in-cloud effect can be included in
models where $f_{nl}\neq 0$. We used the constant barriers as an
example.  We found that the inclusion of the void-in-cloud effect
modifies the abundances of  small voids. As a result, models with
positive $f_{nl}$ show a strong decrement in  very small voids whereas
models with negative $f_{nl}$ show the opposite.  This may due to  the
enhancement of high mass halos for $f_{nl}>0$ which effectively
increases the probability of finding small voids inside a high mass
halo.  Higher order terms in equation~\eqref{eqn:FSint} and a more
accurate description of scale dependent barriers will be needed to
make more quantitative predictions.

Our results have other applications which we have not completed.  
The distribution of the eigenvalues of the initial shear field can 
be used to study the shapes of halos and voids; combining the signals 
with the halo/void abundance and shape distribution would further 
constrain the value of $f_{nl}$.  This is a subtle point because, 
although halo shapes are expected to correlate with the parameters 
$e$ and $p$, we have shown that, at fixed $\delta$, the distribution 
of $e$ and $p$ does not depend on $f_{nl}$. Hence, naively, the shape 
distribution is not informative.  In practice, one usually averages 
over a range of halo masses. In other words, these will have a range of 
$\delta/\sigma$ values, so the result of this averaging may depend 
on $f_{nl}$, for the same reason that the distribution of $e$ 
(equation~\ref{ge}) depends on $f_{nl}$.  
And finally, we are in the process of extending our nonlinear redshift 
space probability distribution function for the dark matter 
\citep{lamshethred} to these $f_{nl}$ models.

\section*{Acknowledgements}
We would like to thank the referee for a helpful report.
V.D. acknowledges support by the Swiss National Foundation under
contract No. 200021-116696/1.


\appendix

\section{Critical value approach}\label{section:halocritical}
The derivation of equation~(\ref{eqn:PS}), which assumes that $F$  is
simply related to $P$, without writing the intermediate steps
associated with the excursion set approach, is ad hoc.   For example,
if we assume the same logic that leads to  equation~(\ref{eqn:PS}) but
choose the integrand  $p(\delta|R)$ so that it returns
equation~(\ref{eqn:ST99}) in the case  $f_{nl}=0$, then we find that
\begin{eqnarray}
\nu_c f(\nu_c) &\equiv& \frac{\partial F}{\partial\ln \nu_c} =
  \frac{\partial}{\partial\,\ln\nu_c} \int^{\infty}_{\sqrt{a}\,\nu_c} d\nu\, 
    2A\left[1 + \frac{\sigma S_3}{6}H_3\right] \,
      \left(1 + \frac{1}{\nu^{2p}}\right)\frac{e^{-\nu^2/2}}{\sqrt{2\pi}}
 \nonumber\\
 &=& \frac{\partial\, F_0^{ST}}{\partial\, \ln\nu_c} \,
  \left(1 + \frac{\sigma S_3}{6}H_3(\sqrt{a}\,\nu_c) \right) \,  
  + \frac{\partial (\sigma S_3/6)}{\partial \ln \nu_c}
       \frac{2A}{\sqrt{2\pi}}  \left\{e^{-a\nu_c^2/2} %
                 \left[a\nu_c^2 - 1 + (a\nu_c^2)^{1-p}\right]  %
  -  \frac{1+2p}{2^p}\,\Gamma\left(1-p,\frac{a\nu_c^2}{2}\right)\right\}.
\label{eqn:STPSra}
\end{eqnarray} 
The first line of this expression has not made its way into the 
$f_{nl}=0$ literature -- a testament to how much more popular the 
excursion set approach (see below) has become.  It does not provide 
a particularly good description of the measured dependence on 
$f_{nl}$, so we do not consider it further.

\section{Bivariate Edgeworth expansion for void abundances}
\label{section:beeva}
We used the bivariate Edgeworth expansion to study 
the first crossing probability when $f_{nl} \neq 0$ in 
\citet{lsfnlhalo}.
We can define the analogy of $\mathcal{G}_{mn}$ here:
\begin{equation}
Q_{mn} = \int^0_{-\infty}{\rm d}\delta\,p_0(\delta+\delta_v,s|\delta_c,S)
              h_{mn}\left(\frac{\delta+\delta_v}{\sqrt{s}},
                 \frac{\delta_c}{\sqrt{S}},\sqrt{\frac{S}{s}}\right),
\end{equation}
where the $h_{mn}$ were defined in \citet{lsfnlhalo},
\begin{align}
Q_{30}  =& \frac{\sqrt{s-S}}{\sqrt{s}}\left[1 
         - \frac{(\delta_v-\delta_c)^2}{s-S}\right]\,
               p_0\left(\frac{\delta_v-\delta_c}{\sqrt{s-S}}\right) 
                                                       \nonumber \\
Q_{03}  = & \frac{(s-S)^2}{s^2}\,H_3(\delta_c/\sqrt{S})
              \frac{1 + {\rm erf}[(\delta_v-\delta_c)/2\sqrt{s-S}]}{2}
                                                         \nonumber \\
 &+   \frac{S(s-S)(2S-3s) + [S^2(\delta_v^2+\delta_v\delta_c+\delta_c^2)-3sS(\delta_v\delta_c+\delta_c^2)+3\delta_c^2s^2]}{s^2\sqrt{S}\sqrt{s-S}}
                                                            \nonumber \\
& \quad \times   p_0\left(\frac{\delta_v-\delta_c}{\sqrt{s-S}}\right) 
                                                            \nonumber \\
Q_{21} = &  -\frac{\sqrt{S}\sqrt{s-S}}{s}
       \left[1 - \frac{(\delta_v-\delta_c)^2}{s-S} 
                  + \frac{\delta_c(\delta_v-\delta_c)}{S}\right]
    p_0\left(\frac{\delta_v-\delta_c}{\sqrt{s-S}}\right) \nonumber \\
Q_{12} =& - \frac{-sS(s-S) + \delta_v^2S^2 -2\delta_v\delta_csS + \delta_c^2s^2}{Ss^{3/2}\sqrt{s-S}}
   p_0\left(\frac{\delta_v-\delta_c}{\sqrt{s-S}}\right),
\end{align}
with $p_0(x) = e^{-x^2/2}/\sqrt{2\pi}$.
In addition, we may define 
\begin{equation}
Q_3\equiv \int_{-\infty}^0 {\rm d}\delta\, 
     p_0(\delta+\delta_v,s|\delta_c,S)H_3(\delta_c/\sqrt{S}) 
 = H_3(\delta_c/\sqrt{S})  
    \frac{1 + {\rm erf}[(\delta_v-\delta_c)/2\sqrt{s-S}]}{2}.
\end{equation}
If we again ignore the scale dependence of $\sigma S_3$, then 
\begin{equation}
f(\delta_v,s|\delta_c,S) =
 2\frac{\partial P(\delta_v,s|\delta_c,S)}{\partial s}
 = f_0(\delta_v,s|\delta_c,S)
    -2 \frac{\sigma S_3}{6}\frac{\partial }{\partial s}
 \left[\mathcal{E}(s,S)p_0\left(\frac{\delta_v-\delta_c}{\sqrt{s-S}}\right)\right],
\end{equation}
where $\mathcal{E}(s,S)$ is the same as in \citet{lsfnlhalo} with 
$b=\delta_v$ and $B=\delta_c$. 

\label{lastpage}
\end{document}